\documentclass[a4paper,11pt]{article}

\usepackage[dvips]{graphicx}

\usepackage{amssymb}
\usepackage{amsmath}
\usepackage{longtable}

\usepackage{cite}

\title{Searches for Magnetic Monopoles and...beyond}

\author{\small G. Giacomelli$^{1,2}$, L. Patrizii$^{1,*}$ and Z. Sahnoun$^{1,3}$}

\date{\small $^{1}$INFN sez. Bologna, v.le Berti Pichat 6/2, I-40127 Bologna, Italy.\\
$^{2}$Phys. Dept., Bologna University, v.le Berti Pichat 6/2, I-40127 Bologna, Italy.\\
$^{3}$Astrophys. Dept. CRAAG, B.P. 63, Bouzareah, Algiers, Algeria.\\
$^*$E-mail: patrizii@bo.infn.it\\
\footnotesize Invited paper at the 5$^{th}$ International Conference on ``Beyond the Standard Models of Particle Physics, Cosmology and Astrophysics'', Cape Town, South Africa, 2010. \\
}

\begin{document}
\maketitle

\begin{abstract}
The searches for classical Magnetic Monopoles (MMs) at accelerators, for GUT Superheavy MMs in the penetrating cosmic radiation and for Intermediate Mass MMs at high altitudes are discussed. The status of the search for other massive exotic particles such as nuclearites and Q-balls is briefly reviewed.
\end{abstract}



\section{Introduction}\label{aba:sec1}
Magnetic Monopoles (MMs) are hypothetical particles carrying a magnetic charge which is quantized according to the Dirac relation~\cite{dirac}: $e \,g=n\hbar c/2= n\times g_{D}$, where $e$ is the basic electric charge, $n=1,2,...$; $g_{D}=\hbar c/2e = 68.5\,e$ is the unit Dirac charge.\par
Pointlike, magnetic charged particles are usually referred to as ``classical" or ``Dirac" monopoles, whose properties are derived from the Dirac relation. No predictions exist for their mass (a rough estimate obtained assuming that the classical monopole radius is equal to the classical electron radius yields $m_{M} \simeq \frac{\displaystyle{g^{2}m_{e}}}{\displaystyle{e^{2}}} \simeq n \ 4700\  m_{e} \simeq n \ 2.4\ $ GeV/c$^{2}$). Dirac MMs have been searched for at every new accelerator/collider.\par
So-called ``primordial'' GUT magnetic monopoles are topological point defects possibly produced in the Early Universe at the phase transition corresponding to the spontaneous breaking of the Unified Gauge group into subgroups, one of which is U(1) \cite{thooft, polyakov}. GUT MMs would have masses as large as $10^{16} - 10^{17}$ GeV$/c^{2}$.\par Later phase transitions could have lead to Intermediate Mass Monopoles (IMMs) \cite{lazaride} with masses in the range $10^{5} \div 10^{13}$ GeV$/c^{2}$.\par

Given their large expected mass, GUT and Intermediate Mass monopoles can only be searched for as relic particles from the Early Universe in the Cosmic Radiation (CR).\par
In this paper we review the experimental situation on MM searches; a short discussion on the searches for nuclearites~\cite{nucleariti} and Q-balls~\cite{qballs} is also presented.

\section{Magnetic Monopole Energy Losses} 
A fast MM with magnetic charge $g_{D}$ and velocity $v=\beta c$ behaves like an equivalent electric charge $(ze)_{eq}=g_{D}\beta$ losing energy mainly by ionization; for $\beta>10^{-1}$, the  energy loss of a $g_{D}$ MM is $\sim (68.5)^{2} \sim 4700$ times that of a minimum ionizing particle.\par
Slow poles ($10^{-4}<\beta<10^{-2}$) lose energy by ionization or excitation of atoms and molecules of the medium (``electronic'' energy loss) or by yielding kinetic energy to recoiling atoms or nuclei (``atomic'' or ``nuclear'' energy loss). Electronic energy loss dominates for $\beta>10^{-3}$. In noble gases and for monopoles with $10^{-4}<\beta<10^{-3}$ there is an additional energy loss due to atomic energy level mixing and crossing (Drell effect \cite{Drell}). \par
At very low velocities ($v<10^{-4}c$) MMs may lose energy in elastic collisions with atoms or with nuclei. The energy is released to the medium in the form of elastic vibrations and/or infrared radiation~\cite{derkaoui1}.\par
\begin{figure}[h]
	\begin{center}
		\includegraphics[width=0.8\textwidth]{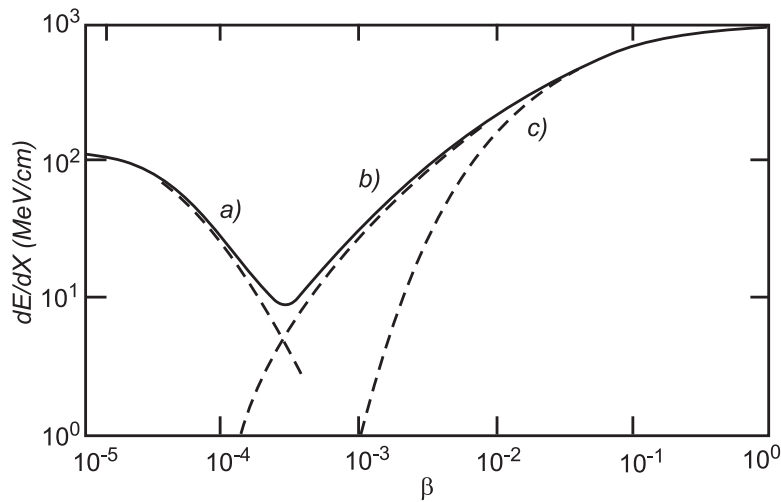}
	\end{center}
	\caption{The energy losses, in MeV/cm, of $g=g_{D}$ MM in liquid hydrogen vs ${\beta}$. Curve a) corresponds to elastic monopole--hydrogen atom scattering; curve b) to interactions with energy level crossings; curve c) sketches the ionization energy loss.}
	\label{fig:perdita-di-energia}
\end{figure}

In Fig. \ref{fig:perdita-di-energia} the different energy loss mechanisms at work in liquid hydrogen for a $g=g_{D}$ MM versus its $\beta$ are shown~\cite{gg+lp}.

\subsection{Searches for Classical (Dirac) Monopoles}
Dirac magnetic monopoles have been searched for at accelerators and colliders in $e^{+}e^{-}$, $e^{+}p$, $p\overline{p}$ and $pp$ collisions, mostly using scintillation counters, wire chambers and nuclear track detectors (NTDs). Searches based on induction devices looking at persistent currents induced by monopoles in superconducting coils were also made.\par
 In Table \ref{tab:CLtable} the accelerator searches for Dirac MMs taken from ref. \cite{pdg} are listed. The most recent searches are briefly discussed in the following sections.

\makeatletter
\setlength{\abovecaptionskip}{6pt}   
\setlength{\belowcaptionskip}{6pt}   

\begin{table}
\caption{List of accelerator based MM searches.}
\centering
\resizebox{1.10\textwidth }{!}
{\begin{tabular} {|l|l|l|l|l|l|l|l|l|}
\hline
Reac-     & $\sqrt{s}$  & Mass range    & Magnetic charge    & Cross Section & Accelerator & Experiment & Year \\
 tion     & (GeV)       & (GeV/c$^{2}$) & (in $g_{D}$ units) & Upper Limit   &             &            & 	 \\
          &             &               &                    & (pb)          & 	           &            & 	 \\ [0.5ex]
\hline\hline
$e^{+}e^{-}$    & 206   & 45-102        & 1       & $0.05$       & LEP2 & OPAL & \'\,08 \\ 
$e^{+}e^{-}$    & 88-94 & $<45; 41.6$   & 1; 2    & $0.3 $       & LEP & MODAL & \'\,93 \\ 
$e^{+}e^{-}$    & 89-93 & $<44.9$       & 1       & $70  $       & LEP & MODAL & \'\,92 \\ 
$e^{+}e^{-}$    & 50-61 & $<29; 18$     & 1; 2    & $0.1 $       & KEK  & TRISTAN & \'\,89 \\ 
$e^{+}e^{-}$    & 50-52 & $<24$         & 1       & $0.8 $       & KEK  & TRISTAN & \'\,88 \\ 
$e^{+}e^{-}$    & 50-52 & $<24$         & 2       & $13$         & KEK  & TRISTAN & \'\,88 \\ 
$e^{+}e^{-}$    & 29    &               & $<3$    & $0.03$       & SLAC & PEP & \'\,84 \\ 
$e^{+}e^{-}$    & 34    & $<10$         & $<6$    & $0.04$       & DESY & PETRA & \'\,83 \\ 
$e^{+}e^{-}$    & 29    & $<30$         & $<3$    & $0.9 $       & SLAC & PEP & \'\,82 \\ [1.0ex]
$e^{+}$p        & 300   &               & 1       & $2   $       & FNAL & HERA & \'\,05 \\
$e^{+}$p        & 300   &               & 2       & $0.2 $       & FNAL & HERA & \'\,05 \\
$e^{+}$p        & 300   &               & 3       & $0.07-0.09$  & FNAL & HERA & \'\,05 \\
$e^{+}$p        & 300   &               & $\ge6$  & $0.05-0.06$  & FNAL & HERA & \'\,05 \\ [1.0ex]
$p\overline{p}$ & 1960  & $200-700$     & 1       & $0.2$        & FNAL & CDF & \'\,06 \\
$p\overline{p}$ & 1800  & $>265$        & 1       & $0.6$        & FNAL & D0 & \'\,04 \\
$p\overline{p}$ & 1800  & $>355$        & 2       & $0.2$        & FNAL & D0 & \'\,04 \\
$p\overline{p}$ & 1800  & $>410$        & 3       & $0.07$       & FNAL & D0 & \'\,04 \\
$p\overline{p}$ & 1800  & $>375$        & 6       & $0.2$        & FNAL & D0 & \'\,04 \\
$p\overline{p}$& 1800  & $>295$         & 1       & $0.7$        & FNAL & FNAL E882 & \'\,00 \\
$p\overline{p}$& 1800  & $>260$         & 2       & $7.8$        & FNAL & FNAL E882 & \'\,00 \\
$p\overline{p}$ & 1800  & $>325$        & 3       & $2.3$        & FNAL & FNAL E882 & \'\,00 \\
$p\overline{p}$ & 1800  & $>420$        & 6       & $0.11$       & FNAL & FNAL E882 & \'\,00 \\
$p\overline{p}$ & 1800  & $<800$        & $\ge1$  & $1.2\times10^{3}$     & FNAL & - & \'\,90 \\
$p\overline{p}$ & 1800  & $<800$        & $\ge1$  & $3\times10^{4}$       & FNAL & - & \'\,87 \\
$p\overline{p}$ & 540   &               & $1,3$   & $1\times10^{5}$       & CERN-ISR &\cite{gg-isr} & \'\,83 \\ [1.0ex]
$pp$            & 52    & $<20$         &         & $8$          & CERN-ISR & \cite{gg-isr}  & \'\,82 \\
$pp$            & 56    & $<30$         & $<3$    & $0.1 $       & CERN-ISR &\cite{gg-isr}  & \'\,78 \\
$pp$            & 63    & $<20$         & $<24$   & $0.1 $       & CERN & \cite{gg-isr}  & \'\,78 \\
$pp$            & 60    & $<30$         & $<3$    & $2$          & CERN-ISR  &\cite{gg-isr}  & \'\,75 \\ [1.0ex]
$pA$            & 11.9  & $<5$          & $<2$    & $1\times10^{-4}$       & IHEP & IHEP & \'\,76  \\
$pA$            &       &               &         & $4\times10^{3}$       & BNL  & - & \'\,76  \\
$pA$            & 28.3  & $<12$         & $<10$   & $5\times10^{-7}$       & FNAL & - & \'\,75  \\
$pA$            & 28.3  & $<13$         & $<24$   & $5\times10^{-6}$       & FNAL & - & \'\,74  \\
$pA$            & 11.9  & $<5$          &         & $1\times10^{-5}$       & IHEP & IHEP & \'\,72  \\
$pA$            & 7.6   & $<3$          & $<2$    & $1\times10^{-4}$       & CERN & - & \'\,63  \\
$pA$            & 7.86  & $<3$          & $<2$    & $2\times10^{-4}$       & BNL-AGS  & - & \'\,63  \\
$pA$            & 7.6   & $<3$          & $<4$    & $10$          & CERN & - & \'\,61  \\
$pA$            & 3.76  & $<1$          & $1$     & $20$          & LBL-Bevatron  & - & \'\,59  \\ [1.0ex]
$nA$            & 24.5  &               &         & $2\times10^{6}$       & FNAL & - & \'\,75  \\ [1.0ex]
$AuAu$          & 4.87  & $<3.3$        & $\ge2$  & $0.65\times10^{3}$    & BNL-AGS  &  -  & \'\,97 \\
$PbA$           & 17.9  & $<8.1$        & $\ge2$  & $1.9\times10^{3}$     & CERN &   -   & \'\,97 \\ 
\hline
\end{tabular}}
\label{tab:CLtable}
\end{table}

\subsubsection{Searches at LEP}
Searches at the CERN LEP $e^{+}e^{-}$ collider were performed by the MODAL \cite{modal} and OPAL \cite{OPAL} collaborations. Both searches were based on the detection of MM pair produced through the $e^{+}e^{-} \rightarrow \gamma^{*} \rightarrow M \overline{M}$ reaction.\par
 The MODAL \cite{modal} experiment was run at $\sqrt{s} = 91.1$ GeV. The detector consisted of a polyhedral array of CR39 NTD foils covering a $0.86 \times 4\pi$ sr angle surrounding the I5 interaction point at LEP. The integrated luminosity was $\sim 60$ nb$^{-1}$. After chemical etching NTD sheets were analysed in the search for penetrating tracks consistent with the passage of a heavily ionizing particle. No candidate event was found; the 95\% CL upper limit on the MM production cross section was $70$ pb for monopoles with masses $<45$ GeV/c$^{2}$.\par
\begin{figure*}
\begin{center}
\resizebox{!}{5.6cm}{\includegraphics{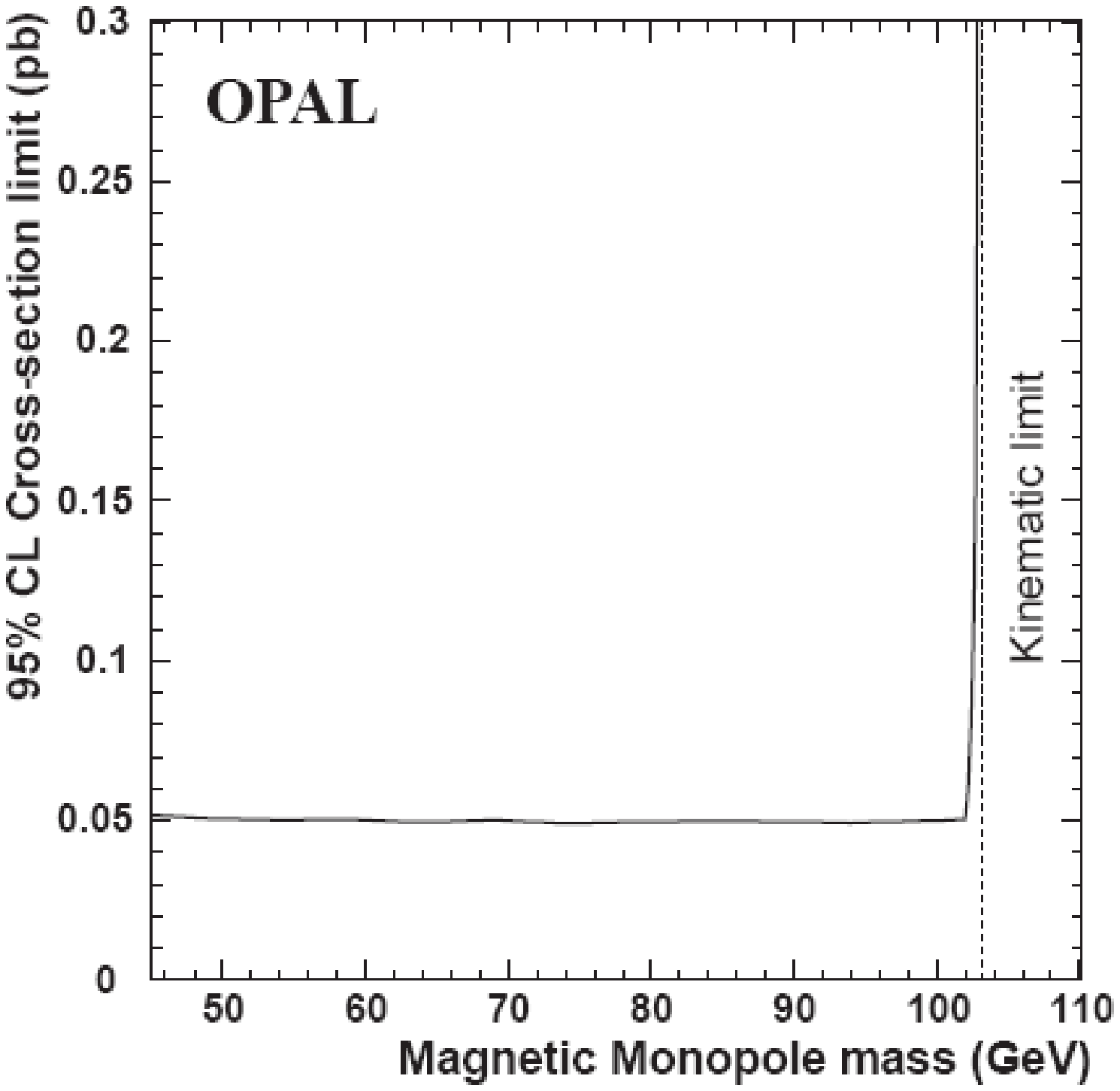}}
\hspace{0.05cm}
\resizebox{!}{5.8cm}{\includegraphics{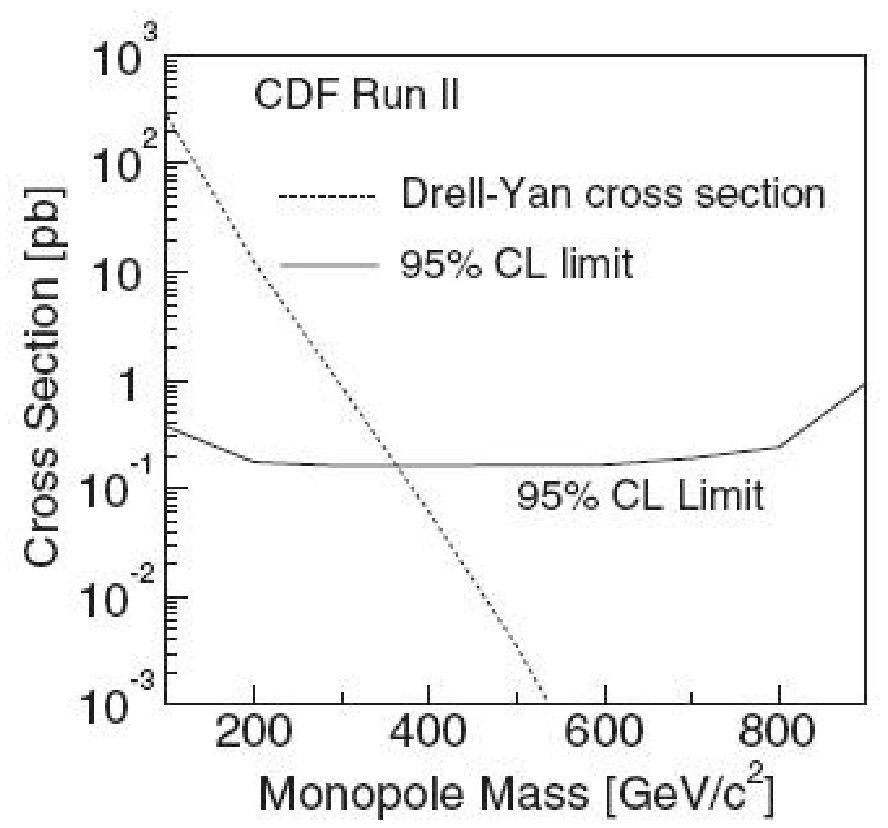}}
\caption{ The 95\% CL upper limits on monopole pair-production cross section versus magnetic monopole mass as obtained by the OPAL (a) and CDF (b) experiments. In Fig.\ref{fig:OPAL}b the Drell-Yan cross section theoretical curve is also given.}
\label{fig:OPAL}
\end{center}
\end{figure*}

The OPAL Collaboration performed a search based on the detection of pair produced MMs at $\sqrt{s}=206.3$ GeV and a total integrated luminosity of $62.7$ pb$^{-1}$  \cite{OPAL}. The search was based on the measurements of the momentum and energy loss in the tracking chambers of the OPAL detector. Back-to-back tracks with high energy release were searched for in opposite sectors of the Jet Chamber. The 95\% CL cross section upper limit for the production of monopoles with masses $45$ GeV/c$^{2}$ $<m_{M}<104$ GeV/c$^{2}$ was $0.05$ pb (Fig. \ref{fig:OPAL}a).

\subsubsection{Searches at HERA}
\begin{figure}
	\begin{center}
		\includegraphics[angle=270, width=1\textwidth]{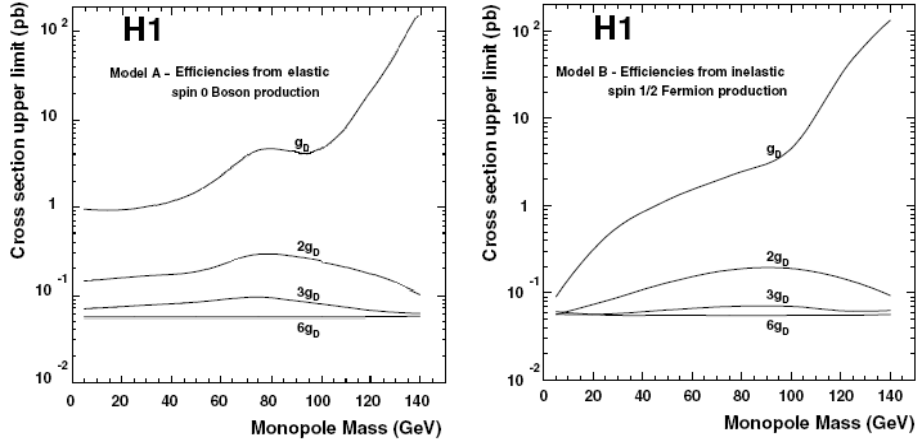}
	\end{center}
	\caption{The cross section upper limits derived from the H1 run of the HERA experiment. Curve a) corresponds to spin 0 MM pair production; curve b) to spin 1/2 MM pair production.}
	\label{fig:HERA}
\end{figure}

The H1 collaboration performed an indirect search for monopoles produced in high energy $e^{+}p$ collisions at $\sqrt{s} = 300$ GeV \cite{tevatron}.  MMs would stop and be trapped in the beam pipe surrounding the H1 interaction point at HERA. The aluminium beam pipe had been exposed to a luminosity of $62\pm1$ pb$^{-1}$ at $\sqrt{s} = 300$ GeV; during HERA operations it was immersed in a $1.15$ T magnetic field parallel to the beam pipe. The beam pipe was cut into long thin strips which were passed through a superconducting coil coupled to a SQUID; the signature of the presence of a MM would be the induction of a persistent current in the superconducting loop. 
Two models for $\mathrm {M \overline{M}} $ pair production were considered: 1) spin 0 monopole pair production by the elastic process $e^{+}p \rightarrow e^{+}p\, M\overline{M} $; 2) spin 1/2 monopole pair production by the inelastic process $e^{+}p \rightarrow e^{+}X\, M\overline{M}$ (where $X$ is any state).
The upper limits on the production cross sections derived for these models are shown in Figure \ref{fig:HERA}.

\subsubsection{Searches at FNAL}

Several searches for MMs were performed at the Tevatron-FNAL $\overline{p}p$ collider.\par
The CDF collaboration in 2006 \cite{CDF} performed a search for magnetic monopoles produced in 35.7 pb$^{-1}$ integrated luminosity of $\overline{p}p$ collisions at $\sqrt{s}= 1.96$ TeV. MMs would have been detected by  the Central Outer Tracker and ToF detectors placed in the 1.4 T magnetic field parallel to the beam direction. The $\mathrm {M \overline{M}}$ pair production was excluded at the 95\% CL for cross sections $< 0.2$ pb and monopole masses in the range $200<m_{M}< 700$ GeV/c$^{2}$ (Fig. \ref{fig:OPAL}b).

 The effects of virtual MMs were looked for searching for $\gamma \gamma$ production via a virtual monopole loop in $p \overline{p}$ collisions at the Tevatron collider. The $\overline{p}p \rightarrow \gamma \gamma$ cross section at energies below the monopole production threshold would be enhanced by the strong coupling of virtual monopoles to photons \cite{loop}.\par

A different indirect search was made looking for monopoles trapped in beam pipes and detector materials from the old D0 and CDF detectors. Several Be, Pb and Al samples were passed through the strong field generated by a superconducting magnet. Trapped monopoles would induce a persistent current in the coil after the passage of the samples\cite{Milton}. This technique, which is independent of the magnetic monopole mass and velocity, was used also in the search for cosmic MMs in bulk matter by passing through the superconducting magnet moon rocks, meteorites, schists and terrestrial magnetic materials \cite{gg1}.

\subsubsection{MoEDAL: Monopole Searches at the LHC}
\begin{figure}
	\begin{center}
		\includegraphics[angle=270, width=0.8\textwidth]{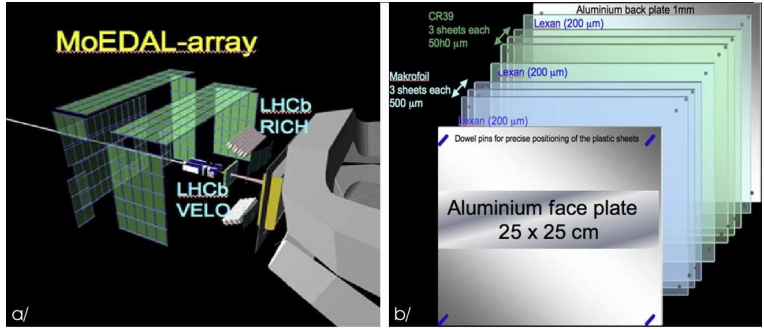}
	\end{center}
	\caption{a) Skecth of the MoEDAL detector as planned to be deployed in the LHCb VELO region. b) A MoEDAL detector element consisting of sheets of CR39, Makrofol and Lexan NTDs.}
	\label{fig:moedal}
\end{figure}

MoEDAL (Monopole and Exotic particle Detection At the LHC) is a future  experiment at the LHC \cite{Moedal}. It will search for MMs and other highly ionizing exotic particles in $p\,p$ collisions at an expected luminosity of $10^{32}$ cm$^{-2}$ s$^{-1}$ and also in the heavy ion running. 
The MoEDAL detector will be an array of NTD stacks deployed around the Point-8 intersection region of the LHCb detector, in the VELO cavern as sketched in Fig. \ref{fig:moedal}a. The array will cover a surface area of $\sim 25$ m$^{2}$. Each stack, $25 \times 25$ cm$^{2}$, will consist of 9 interleaved layers of CR39, Makrofol and Lexan NTDs (Fig. \ref{fig:moedal}b) .

\begin{figure} [t]
	\begin{center}
		\includegraphics[width=0.88\textwidth]{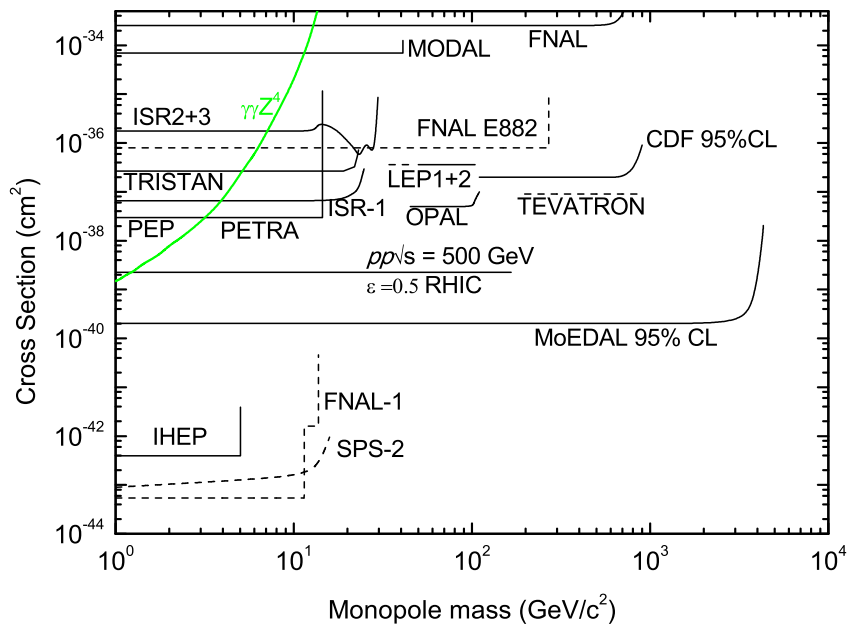}
	\end{center}
	\caption{Cross section upper limits on monopole production from past searches at accelerators/colliders. Dashed lines indicate limits from indirect searches. The upper limit expected from the planned MoEDAL experiment at LHC is also indicated.}
	\label{fig:MMUpLim}
\end{figure}

The passage of a heavily ionizing particle in NTD foils would cause the formation of a damage (latent-track) along its trajectory; a subsequent chemical etching would lead to the formation of etch-pit cones in both front and back faces of each sheet. The size and shape of the cones are related to the particle restricted energy loss and angle of incidence. A detailed description of NTD  technique can be found in ref. \cite{slim}. The MM signature in MoEDAL would be a sequence of collinear etch-pits consistent with the passage of a particle with constant energy loss through the detector foils of a whole stack.\\
\linebreak
In Fig. \ref{fig:MMUpLim} the upper limits on cross sections for magnetic monopole production set by past searches are reported; the expected sensitivity for the LHC-MoEDAL experiment is also shown.

\section{Searches for SuperMassive Magnetic Monopoles}
GUT MMs from the Early Universe may be present today in the cosmic radiation as ``relic'' particles with a velocity spectrum in the $4 \times 10^{-5} <\beta <0.1$ range. Larger velocities could be achieved by IMMs ($10^{5}<m_{M}<10^{13}$ GeV/c$^{2}$) accelerated in one coherent domain of the galactic magnetic field.\par
 Bounds on the flux of cosmic MMs were obtained on the basis of astrophysical and cosmological considerations. The most referred one is the so-called ``Parker Bound'' F$< 10^{-15}$ cm$^{-2}$~s$^{-1}$~sr$^{-1}$ \cite{parker} obtained by requiring that the kinetic energy per unit time that MMs gain from the galactic magnetic field be not larger than the magnetic energy generated in the galaxy by the dynamo effect.
The original limit was re--examined to take into account the almost chaotic nature of the galactic magnetic field, with domain lengths of about $\ell\sim 1$ kpc; the limit becomes  mass dependent \cite{parker}.
 By applying similar considerations to the survival of an early seed of galactic magnetic field a more stringent ``Extended Parker Bound'' (EPB) was obtained \cite{adams}: F $< m_{17}\,10^{-16}$ cm$^{-2}$~s$^{-1}$~sr$^{-1}$, with $m_{17}=m_{M}/10^{17}$ GeV/c$^{2}$. \par
 Several searches for GUT MMs were performed above ground and underground using many types of detectors \cite{ruzicka,gg2,picture}. In Table \ref{tab:GUTtable} are listed the different searches and their results.\par
The most stringent experimental flux limit on supermassive MMs was set by the MACRO experiment at the underground Gran Sasso Laboratory \cite{mm_macro}. 
Searches with underwater and underice neutrino telescopes are sensitive to relativistic MMs. They would be detected by the large amount of Cherenkov radiation, $\gtrsim 8300$ times that of muons.\par
\begin{figure}[ht]
	\begin{center}
		\includegraphics[width=0.9\textwidth]{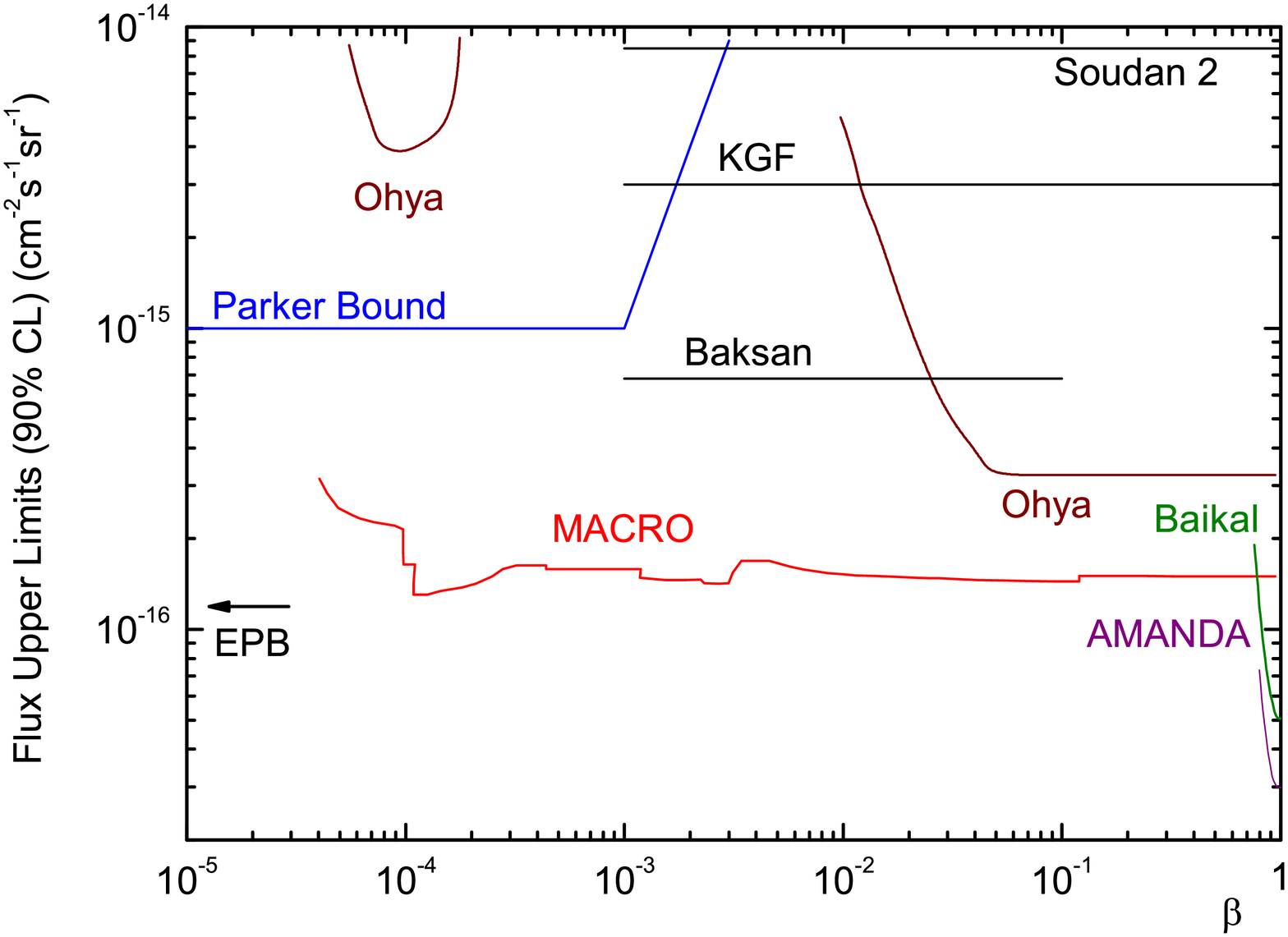}
	\end{center}
	\caption{The 90\% CL upper limits vs $\beta$ for a flux of cosmic GUT monopoles with magnetic charge $g=g_{D}$.}
	\label{fig:GUTlimit}
\end{figure}

 In  Fig.\ref{fig:GUTlimit} the 90\% CL flux upper limits versus $\beta$ for GUT MMs with $g=g_{D}$ as set by the MACRO \cite{mm_macro}, Ohya \cite{ohya}, Baksan \cite{baksan}, Baikal \cite{baikal}, and AMANDA \cite{amanda} experiments are shown. The Baikal and AMANDA limits were obtained assuming that relativistic GUT MMs would reach the detector from ``below'', i.e. after crossing the Earth (which is unlikely).

\begin{table}
\caption{Flux upper limits for GUT and Intermediate Mass Monopoles from different experiments, assuming $g=g_{D}$.}
\centering
\resizebox{1.05\textwidth }{!}
{\begin{tabular} {|l|c|c|c|l|}
\hline
Experiment  &  Mass Range    & $\beta$ range & Flux Upper Limit                & Detection Technique\\
            & (GeV/c${^2})$  &               & (cm$^{-2}$~s$^{-1}$~sr$^{-1}$)  &                     \\ [0.5ex]
\hline
\hline 
AMANDA II Upgoing \cite{amanda} & $10^{11}-10^{14}$ & $0.76-1$ & $8.8-0.38\times 10^{-16}$ & Ice Cherenkov \\
AMANDA II Downgoing \cite{amanda} & $10^{8}-10^{14}$ & $0.8-1$ & $17-2.9\times 10^{-16}$ & Ice Cherenkov\\
AMANDA II (catalysis) \cite{amanda2} & $>10^{11}$ & $\simeq 10^{-3}$ & $5\times 10^{-17}$ & Ice Cherenkov \\[0.5ex]
Baikal \cite{baikal} & $10^{7}-10^{14}$ & $0.8-1$ & $1.83-0.46\times 10^{-16}$ & Water Cherenkov \\
Baikal (catalysis) \cite{baksan} & $5\times10^{13}$ & $ \simeq 10^{-5}$ & $6 \times 10^{-17}$ & Water Cherenkov \\ [0.5ex]
ANTARES \cite{antares} &  $10^{7}-10^{14}$ & $ 0.65-1$ & $9.1-1.3\times 10^{-17}$ & Water Cherenkov \\ [0.5ex]
Super-Kamiokande (catalysis) \cite{SK} & $>10^{17}$ & $10^{-5}-10^{-2}$ & $8\times10^{-27}-3\times10^{-22}$ & Water Cherenkov \\ [0.5ex]
MACRO \cite{mm_macro} & $5\times10^{8}- 5\times10^{13}$ & $ > 5\times10^{-2} $ & $3\times 10^{-16}$ & Scint.+Stream.+NTDs \\
MACRO \cite{mm_macro} & $>5\times10^{13}$ & $ > 4 \times 10^{-5} $ & $1.4\times 10^{-16}$ & Scint.+Stream.+NTDs \\
MACRO (catalysis) \cite{catalisi} & $5\times10^{13}$ & $ > 4 \times 10^{-5} $ & $3-8 \times 10^{-16}$ & Sctreamer tube \\ [0.5ex]
OHYA \cite{ohya} & $5\times10^{7}- 5\times10^{13}$ & $ > 5\times10^{-2} $ & $6.4\times 10^{-16}$ & Plastic NTDs \\
OHYA \cite{ohya} & $>5\times10^{13}$ & $ > 3\times 10^{-2} $ & $3.2\times 10^{-16}$ & Plastic NTDs \\
[0.5ex]
SLIM \cite{slim} & $10^{5}- 5\times10^{13}$ & $ > 3\times10^{-2} $ & $1.3\times 10^{-15}$ & Plastic NTDs \\
SLIM \cite{slim} & $>5\times10^{13}$ & $ > 4 \times 10^{-5} $ & $0.65\times 10^{-15}$  & Plastic NTDs \\[0.5ex]

MICA \cite{price2} & -- & $10^{-4}-10^{-3}$ & $\sim 10^{-17}$ & NTD \\[0.5ex]
INDU Combined \cite{gg+lp,gg1} & $>10^{5}$ & -- & $2 \times 10^{-14}$ & Induction \\
\hline
\end{tabular}}
\label{tab:GUTtable}
\end{table}

The interaction of the GUT monopole core with a nucleon can lead to nucleon decay (catalysis), f.e. \( M + p \rightarrow M + e^{+} + \pi^{0}\) via the Rubakov-Callan mechanism~\cite{rubakov}. Searches for MM induced nucleon  decay were made with neutrino telescopes \cite{amanda2,baksan} and by the MACRO experiment \cite{catalisi}. The Super-Kamiokande collaboration also performed an indirect search for GUT monopoles, looking for a neutrino signal coming from proton decay catalised by GUT MMs captured in the Sun. The flux upper limit was set at $F < 10^{-23}$ cm$^{-2}$~s$^{-1}$~sr$^{-1}$ for a monopole mass $<10^{17}$ GeV/c$^{2}$ and velocity $\beta=10^{-3}$ assuming a catalysis cross section $\sigma\sim1$~mb~\cite{SK}. \\
MMs with masses $m_{M}>10^{5}-10^{6}$ GeV~\cite{derkaoui1} could reach the Earth surface from above and be detected; lower mass MMs may be searched for with detectors located at high mountain altitudes, balloons and satellites.


\begin{figure*}
	\begin{center}
    \resizebox{!}{7.5cm}{\includegraphics{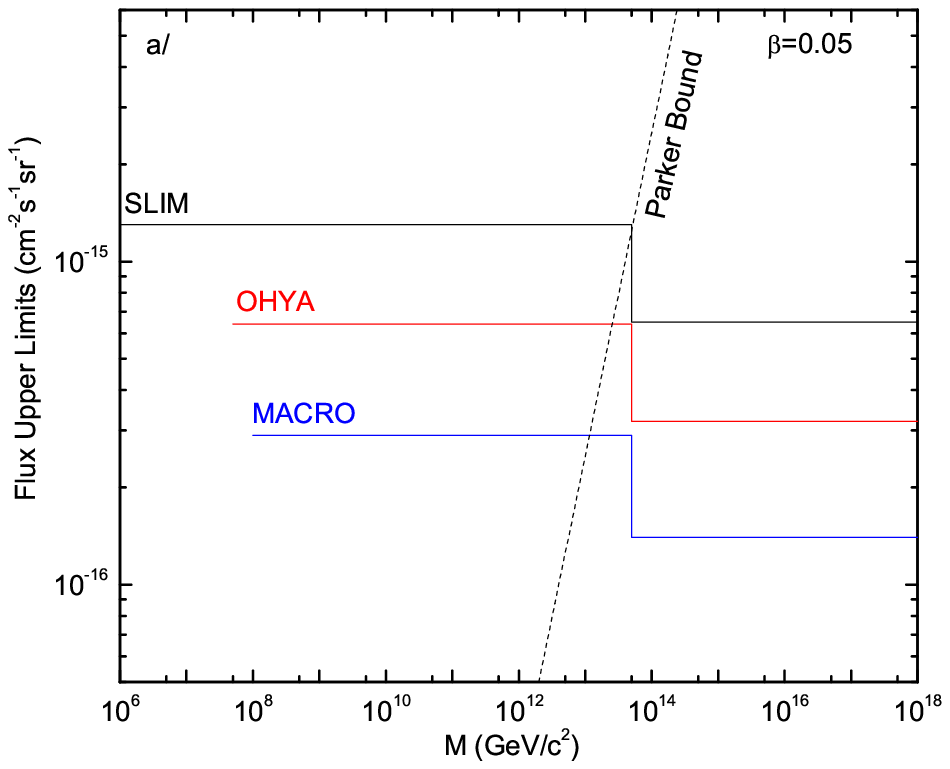}}
    \hspace{0.0cm}
    \resizebox{!}{7.5cm}{\includegraphics{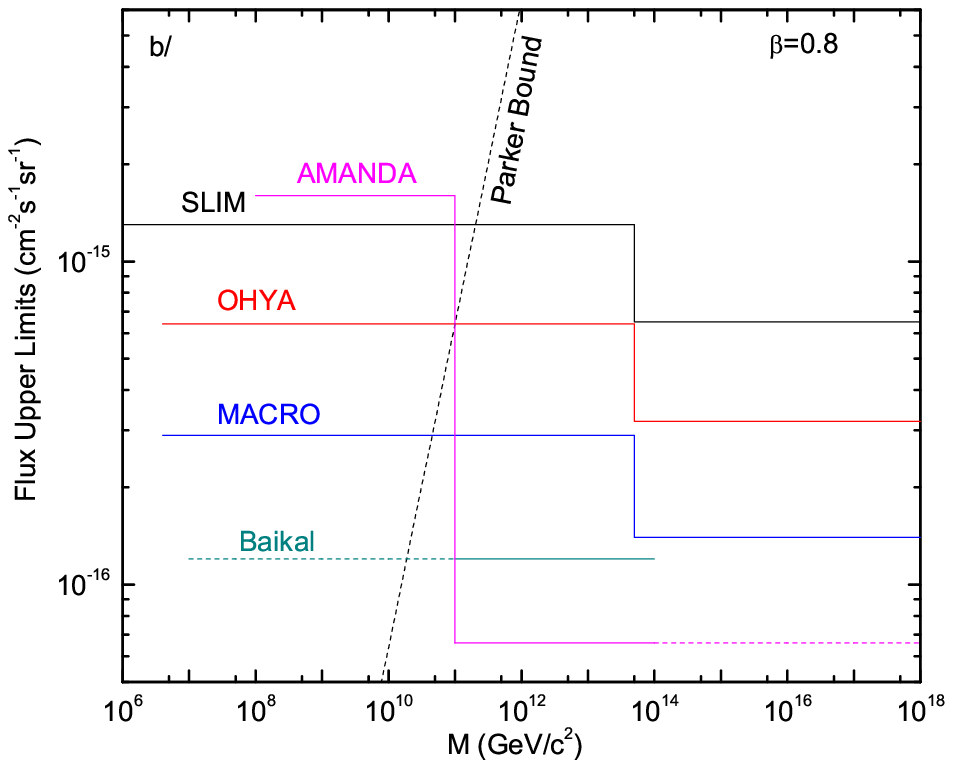}}
	\caption{Experimental 90\% CL flux upper limits versus mass for MMs with (a) $\beta = 0.05$, (b) $\beta = 0.8$ at the detector level from different experiments.}
	\label{fig:monopoles}
	\end{center}
\end{figure*}

The SLIM experiment at the Chacaltaya high altitude laboratory (5230 m a.s.l.)~\cite{slim} searched for downgoing IMMs with a 427 m$^{2}$ NTD array exposed for 4.2 years to the cosmic radiation. SLIM was sensitive to MMs with $g=2g_{D}$ in the whole range $4 \times 10^{-5}<\beta <1$ and $g=g_{D}$ for $\beta >10^{-3}$. No candidate event was observed. 

Flux upper limits versus mass, as set by SLIM and other experiments for two different MM velocities, are plotted in Fig.\ref{fig:monopoles}. \par

Constraints on the flux of ultra-relativistic MMs were also given by two experiments based on the detection of radio wave pulses from the interaction of a primary particle with matter (ice). The Radio Ice Cherenkov Experiment, RICE, consisting of radio antennas buried in the Antarctic ice sets a flux upper limit of the order of $10^{-18}$ cm$^{-2}$~s$^{-1}$~sr$^{-1}$ for intermediate-mass monopoles with a Lorentz factor $10^{7}<\gamma<10^{12}$ and an anticipated total energy of E=$10^{16}$ GeV (the monopole rest mass being E/$\gamma$)~\cite{rice}. The ANITA-II Balloon-borne radio Interferometer determined a 90$\%$ CL flux upper limit of the order of $10^{-19}$ cm$^{-2}$~s$^{-1}$~sr$^{-1}$ for $\gamma>10^{10}$ at the total energy of $10^{16}$ GeV \cite{anita}. \par

\section{Searches for Strange Quark Matter and Q-balls}
Strange Quark Matter (SQM) composed of approximately the same number of up, down and strange quarks was conjectured as the ground state of nuclear matter \cite{nucleariti}. SQM density would be larger than that of atomic nuclei and be stable for all baryon numbers in the range $300<A<10^{57}$. Due to the suppression of some \textit{s} quarks SQM should have a relatively small positive electric charge \cite{Heiselberg, Madsen00} neutralized by an electron cloud, thus forming a sort of atom. Large lumps of SQM ($A>10^{10}$) can be present in the cosmic radiation (``nuclearites''). They could have been produced in the early Universe and be a component of the galactic cold dark matter with typical velocities of $\sim$ $10^{-3}$c~\cite{nucleariti}. 
The main energy loss mechanism for galactic nuclearites is that of elastic or quasi-elastic collisions with the ambient atoms and molecules. The energy loss is large; nuclearites should be easily detected by scintillators and NTDs~\cite{macro-nucl}. In transparent media some of the energy dissipated could appear as visible light.\par
\begin{figure}[ht]
\begin{center}
\includegraphics[width=0.72\textwidth]{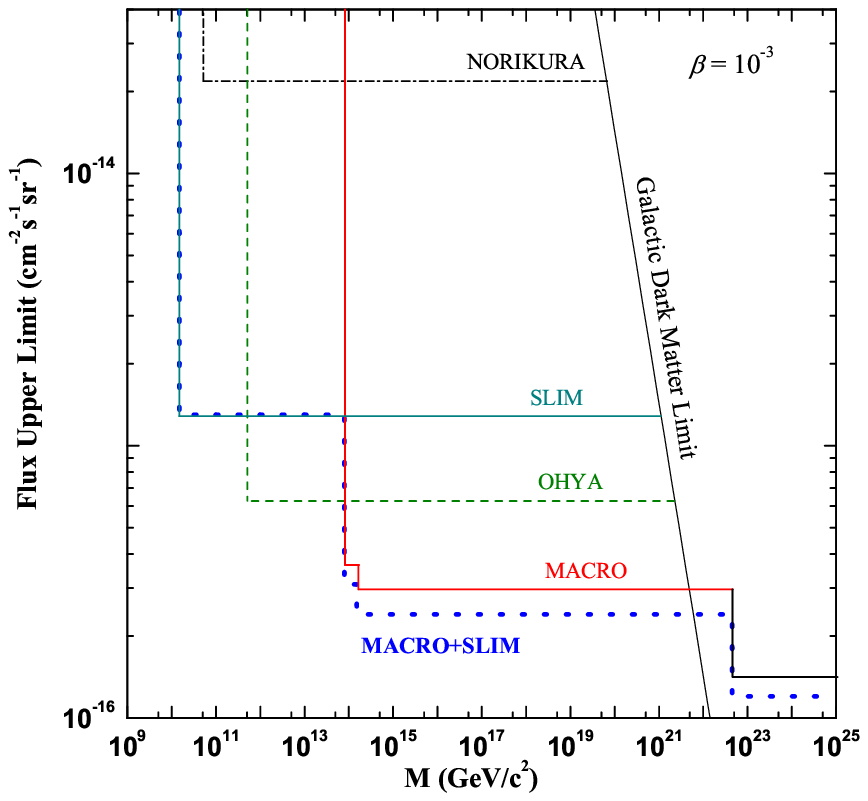}
\end{center}
\caption{90\% CL flux upper limits vs mass for intermediate and high mass nuclearites with $\beta=10^{-3}$ obtained from various searches with NTDs. A combined flux from the MACRO and SLIM experiments is also shown.}
\label{fig:NuclLimitComp}
\end{figure}

 Several flux upper limits on nuclearites were obtained as by-products of magnetic monopole searches. In Fig.~\ref{fig:NuclLimitComp} the most stringent limits for nuclearites with $\beta=10^{-3}$ are at the level of $1.4 \div 3 \times 10^{-16}$ cm$^{-2}$s$^{-1}$sr$^{-1}$ \cite{SQM}. In the figure the galactic dark matter bound is obtained assuming that all dark matter is composed of nuclearites.

Smaller SQM bags with baryon number $A<10^{6}\div10^{7}$ are usually called ``strangelets''. They could be produced in very energetic astrophysical processes involving strange star collisions~\cite{Madsen88,Friedman} and supernovae explosions~\cite{Vucettich}.\par
\begin{figure}[ht]
\begin{center}
\includegraphics[width=0.72\textwidth]{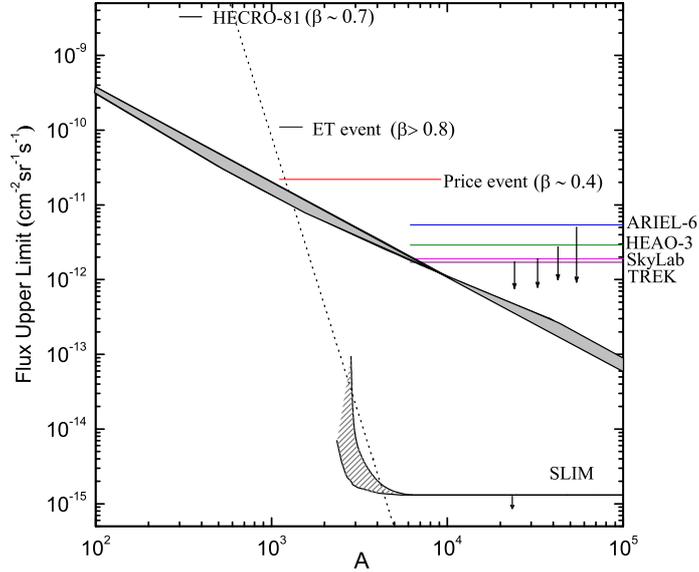}
\end{center}
\caption{90\% CL flux upper limits vs mass number A for relativistic strangelets from onboard balloon and space experiments and by the SLIM detector at mountain altitude. The grey band and the dotted line are the fluxes predicted by different models \cite{Madsen05,WilkF}.} 

\label{fig:StrLimitComp}
\end{figure}

Strangelets are generally assumed to have no associated electrons; their interaction with matter should be similar to that of heavy ions  with a different charge to mass ratio, $Z/A$, and they would undergo the same acceleration and interaction processes as ordinary cosmic rays. In Fig.~\ref{fig:StrLimitComp} the flux upper limits for relativistic strangelets obtained with the SLIM detector and by experiments onboard stratospheric balloons and in space are given \cite{SQM,Ariel,Heao,Skylab,Trek}. The three uppermost horizontal lines in the figure indicate the measured flux assuming that unusual events found in  cosmic rays could be due to SQM \cite{price2,Hecro,ET,AMS2}. The fluxes expected according to different models for SQM propagation in the Galaxy and in the atmosphere are indicated in Fig.~\ref{fig:StrLimitComp} by the dotted line and the grey band \cite{Madsen05,WilkF}. Limits for small mass strangelets could come from satellite experiments as for example AMS-2 on the International Space Station which will have an estimated sensitivity at the level of $\sim10^{-12}$ cm$^{-2}$s$^{-1}$sr$^{-1}$ \cite{ams}.\\

\begin{figure*}
\begin{center}
         \includegraphics[width=0.7\textwidth]{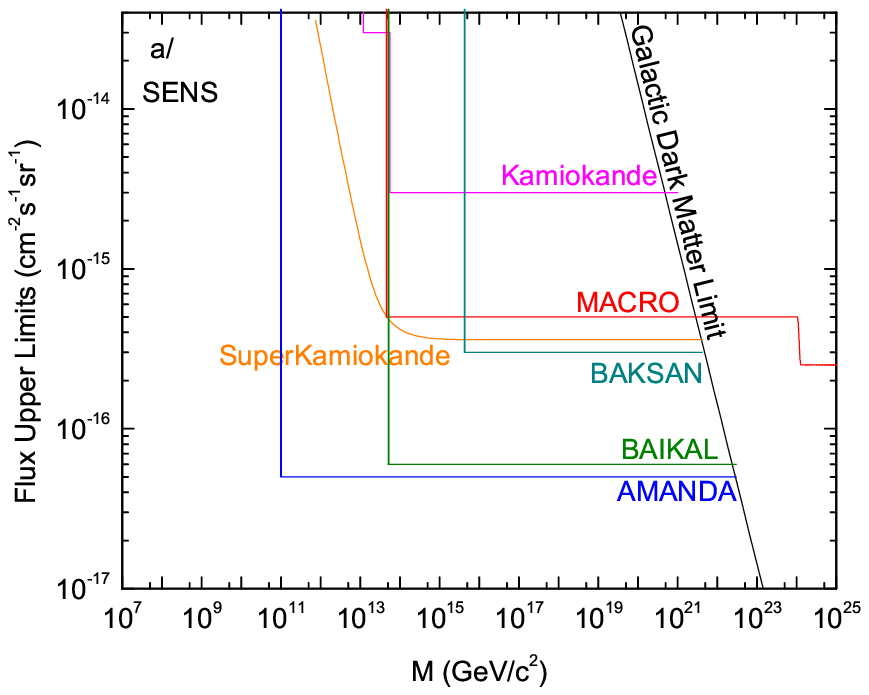}
\hspace{0.0cm}
         \includegraphics[width=0.7\textwidth]{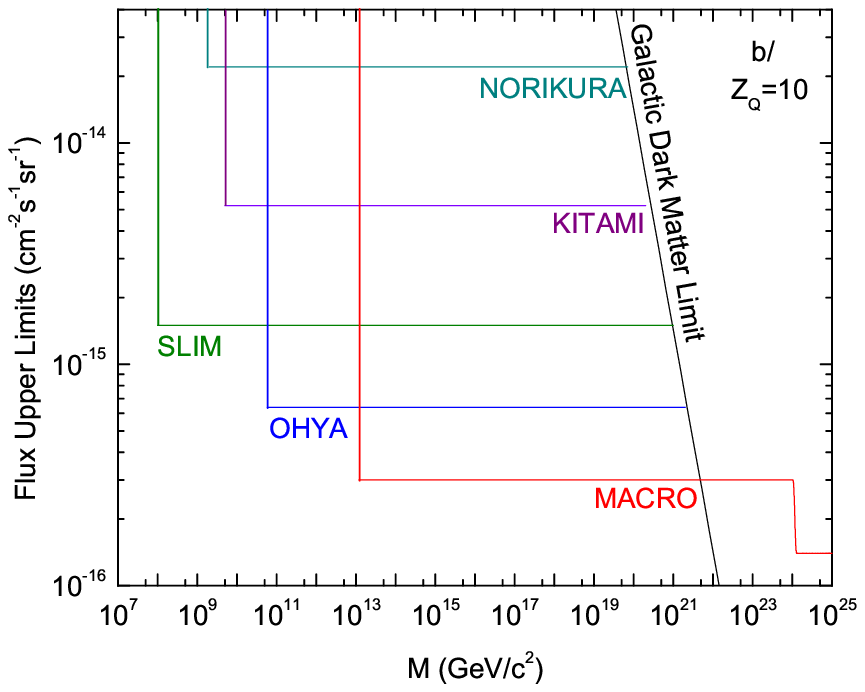}
\caption{Flux upper limits versus mass for: a/ Neutral Q-balls and b/ charged Q-balls with $Z_{Q} >10e$ obtained by various experiments.}
\label{fig:QballLimitComp}
\end{center}
\end{figure*}

Q-balls are hypothesized coherent states of squarks $\tilde{q}$, sleptons $\tilde{l}$ and Higgs fields predicted by minimal supersymmetric generalizations of the Standard Model of particle physics~\cite{qballs}. They may carry some conserved global baryonic charge Q and possibly also a lepton number. Q-balls could have been copiously produced in the early Universe and may have survived till now as a dark matter component. They are classified into two groups $(i)$ neutral Q-balls, generally called SENS (Supersymmetric Electrically Neutral Solitons) that should be massive and may catalyse proton decay and $(ii)$ charged Q-balls called SECS (Supersymmetric Electrically Charged Solitons) that might be formed by SENS gaining an integer electric charge from proton or nuclei absorption. 
SECS with typical galactic velocities $\beta \simeq 10^{-3}$ and $M_{Q} < 10^{13}$ GeV could reach an underground detector from above, SENS also from below. SENS may be detected by their continuons emission of charged pions (energy loss $\sim$100 GeV g$^{-1}$cm$^2$), generally in large neutrino telescopes. SECS may interact in a way not too different from nuclearites and may be detected by scintillators, NTDs and ionization detectors. In Fig.~\ref{fig:QballLimitComp} are shown the flux upper limits versus mass for SENS and for charged Q-balls (with $Z_{Q}>$ $10 e$) set by various experiments. Note that most of these limits were given from re-estimates of the experimental limits set for MM searches (f.e. \cite{Arafune}). The SENS upper limit given by Super-Kamiokande in a dedicated analysis is also shown \cite{SK2}. Not shown in the figure is the result obtained  by the DAMA Collaboration which set an upper limit on the flux of charged Q-balls with $\beta\simeq 10^{-3}$ at the level of $\sim 3\times 10^{-11}$ cm$^{-2}$s$^{-1}$sr$^{-1}$ \cite{dama}.

\section{Conclusions}
Searches for classical Dirac MMs at accelerators set limits on the pair production cross sections from different physical processes and for a wide range of masses, as shown in Fig.~\ref{fig:MMUpLim}. Future improvements may come from new experiments at the LHC~\cite{Moedal}.\par
Many searches were performed for GUT monopoles in the penetrating cosmic radiation. The 90\% CL flux limits are at the level of $\sim1.4 \times 10^{-16} $~cm$^{-2}$~s$^{-1}$~sr$^{-1}$ for $\beta \ge 4 \times 10^{-5}$. It may be difficult to do much better unless new refined detectors with considerably larger areas are proposed.\par
Present limits on Intermediate Mass Monopoles with high $\beta$ are at the level of $\sim1.3 \times 10^{-15} $~cm$^{-2}$~s$^{-1}$~sr$^{-1}$ given by experiments at high altitudes. These limits could be improved with much larger detectors, in particular by large volume neutrino telescopes. IceCube expects limits of the order of $\sim10^{-17}$~cm$^{-2}$~s$^{-1}$~sr$^{-1}$ for GUT MMs inducing nucleon catalysis  and $\sim10^{-18} - 10^{-19}$~cm$^{-2}$~s$^{-1}$~sr$^{-1}$ for relativistic poles \cite{icecube}.\par
As a by-product of GUT MM searches some experiments obtained stringent limits on nuclearites and on Q-balls fluxes in the cosmic radiation. Future searches with neutrino telescopes \cite{pavalas} and in space \cite{AMS2,ams} should reach sensitivities to nuclearites, strangelets and Q-balls of smaller masses.

\section{Acknowledgments}
Z. S.  thanks INFN, Sez. Bologna for providing FAI Grants for foreigners. We acknowledge the collaboration from many colleagues.


\begin{thebibliography}{9}
\bibitem{dirac} P.A.M. Dirac, {\em Proc. R. Soc. London} {\bf 133}, 60 (1931); {\em Phys. Rev.} {\bf 74}, 817 (1948).
\bibitem{thooft} G.'t Hooft, {\em Nucl. Phys.} {\bf B29}, 276 (1974). 
\bibitem{polyakov} A.M. Polyakov, {\em JETP Lett.} {\bf 20}, 194 (1974); N.S. Craigie et al., Theory and Detection of MMs in Gauge Theories, World Scientific, Singapore (1986).
\bibitem{lazaride} G. Lazarides et al., {\em Phys. Rev. Lett.} {\bf 58}, 1707 (1987); T. W. Kephart and Q. Shafi, {\em Phys. Lett.} {\bf B520}, 313 (2001).
\bibitem{nucleariti} E. Witten, {\em Phys. Rev.} {\bf D30}, 272 (1984); A. De Rujula and S. Glashow, {\em Nature} {\bf 31},272 (1984).
\bibitem{qballs} S. Coleman, {\em Nucl. Phys.} {\bf B262}, 293 (1985); A. Kusenko and A. Shaposhnikov, {\em Phys. Lett.} {\bf B418}, 46 (1998).
\bibitem{Drell} G.F. Drell et al., {\em Nucl. Phys.} {\bf B209}, 45 (1982).
\bibitem{derkaoui1} J. Derkaoui et al., {\em Astrop. Phys.} {\bf 9}, 173 (1998); {\em Astrop. Phys.} {\bf 9}, 339 (1999).
\bibitem{gg+lp} G. Giacomelli et al. hep-ex/011209; hep-ex/0302011; hep-ex/0211035.
\bibitem{pdg} K. Nakamura et al. (Particle Data Group), {\em J. Phys.} {\bf G37}, 075021 (2010) and refs. therein.
\bibitem{gg-isr} For the ISR monopole searches see G. Giacomelli and M. Jacob, {\em Phys. Rept.} {\bf 55}, 1 (1979)
\bibitem{modal} K. Kinoshita et al., {\em Phy. Rev.} {\bf D46}, R881 (1992).
\bibitem{OPAL} G. Abbiendi et al., {\em Phys. Lett.} {\bf B663}, 37 (2008).
\bibitem{tevatron} A. Aktas et al. (The H1 collaboration), {\em Eur. Phys. J.} {\bf C41}, 133 (2005).
\bibitem{CDF} A. Abulencia et al., {\em Phys. Rev. Lett.} {\bf 96}, 201801 (2006).
\bibitem{loop} I.F. Ginzburg and A. Schiller, {\em Phys. Rev.} {\bf D60}, 075016 (1999).
\bibitem{Milton} K.A. Milton, {\em Rept.Prog.Phys.} {\bf 69}, 1637 (2006).
\bibitem{gg1} G. Giacomelli, {\em Riv. Nuovo Cimento} {\bf 7}N12, 1 (1984).
\bibitem{Moedal} J. L. Pinfold et al., {\em Radiat. Meas.} {\bf 44}, 834 (2009).
\bibitem{slim} S. Balestra et al., {\em Eur. Phys. J.} {\bf C55}, 57 (2008).
\bibitem{parker} E.N. Parker, {\em Ap. J.} {\bf 160}, 383 (1970); M.S. Turner et al., {\em Phys. Rev.} {\bf D26}, 1296 (1982).
\bibitem{adams} F.C. Adams et al., {\em Phys. Rev. Lett.} {\bf 70}, 2511 (1993).
\bibitem{ruzicka} J. Ruzicka and V.P. Zrelov, {\em JINR}{\bf-1-2-80-850} (1980).
\bibitem{gg2} G. Giacomelli et al., hep-ex/0005041.
\bibitem{picture} D. Bakari et al., hep-ex/0004019.
\bibitem{amanda} R. Abbasi et al., {\em Eur. Phys. J.} {\bf C69}, 361 (2010); H. Wissing et al., 30$^{th}$ ICRC {\bf 4}, 799 (2007). arXiv:0711.0353 [astro-ph].
\bibitem{amanda2} A. Pohl et al., {\em Proc. International Workshop on Exotic Physics with Neutrino Telescopes}, 77 (2006).
\bibitem{baikal} V. Aynutdinov et al., astro-ph/0507713; R. Wischnewski et al. 30$^{th}$ ICRC, arXiv:0710.3064 [astro-ph].
\bibitem{baksan} E.N. Alexeyev et al., 21$^{st}$ ICRC {\bf 10}, 83 (1990); Yu.F. Novoseltsev et al., {\em Nucl. Phys.} {\bf B151}, 337 (2006).
\bibitem{antares} G. Giacomelli, arXiv:1105.1245 [astro-ph.IM].
\bibitem{SK}  K. Ueno et al., Proc. of the 31$^{st}$ ICRC, Lodz, Poland (2009), http://icrc2009.uni.lodz.pl/proc/pdf/icrc0670.pdf
\bibitem{mm_macro} M. Ambrosio et al., MACRO Coll., {\em Eur. Phys. J.} {\bf C25}, 511 (2002); {\em Phys. Lett.} {\bf B406}, 249 (1997); {\em Phys. Rev. Lett.} {\bf 72}, 608 (1994).
\bibitem{catalisi} M. Ambrosio et al., {\em Eur. Phys. J.} {\bf C26}, 163 (2002).
\bibitem{ohya}  S. Orito et al., {\em Phys. Rev. Lett.} {\bf 66}, 1951 (1991). 
\bibitem{price2} P. B. Price and M. H. Salamon, {\em Phys. Rev. Lett.} {\bf 56}, 1226 (1986); D. Ghosh and S. Chatterjea, {\em Europhys. Lett.} {\bf 12}, 25 (1990).
\bibitem{rubakov} V.A. Rubakov, {\em JETP Lett.} {\bf B219}, 644 (1981); G.G. Callan, {\em Phys. Rev.} {\bf D26}, 2058 (1982).
\bibitem{rice}  D.P. Hogan et al., {\em Phys. Rev.} {\bf D78}, 075031 (2008). 
\bibitem{anita}  M. Detrixhe et al., {\em Phys. Rev.} {\bf D83}, 023513 (2011). 
\bibitem{Heiselberg} H. Heiselberg, {\em Phys. Rev.} \textbf{D48}, 1418 (1993).
\bibitem{Madsen00} J. Madsen, Phys. Rev. Lett. \textbf{85}, 4687 (2000).
\bibitem{Madsen88} J. Madsen, {\em Phys. Rev. Lett.} \textbf{61}, 2909 (1988).
\bibitem{Friedman} J. L. Friedman and R. R. Caldwell, {\em Phys. Lett.} \textbf{B264}, 143 (1991).
\bibitem{Vucettich} H. Vucetich and J. E. Horvath, {\em Phys. Rev.} \textbf{D57}, 5959 (1998).
\bibitem{macro-nucl} M. Ambrosio et al., {\em Eur. Phys. J.} {\bf C13}, 453 (2000).
\bibitem{SQM} S. Cecchini et al., {\em Eur. Phys. J.} {\bf C57}, 525 (2008).
\bibitem{Ariel} P. H. Fowler et al., {\em Astrophys. J.} \textbf{314}, 739 (1987).
\bibitem{Heao} W. R. Binn et al., {\em Astrophys. J.} \textbf{347}, 997 (1989).
\bibitem{Skylab} E. K. Shirk and P. B. Price, {\em Astrophys. J.} \textbf{220}, 719 (1978).
\bibitem{Trek} A. J. Westphal et al., {\em Nature} \textbf{396}, 50 (1998); B. A. Weaver et al., {\em Nucl. Instrum. Meth.} \textbf{B145}, 409 (1998).
\bibitem{Hecro} T. Saito et al., {\em Phys. Rev. Lett.} \textbf{65}, 2094 (1990).
\bibitem{ET} M. Ichimura et al., {\em Il Nuovo Cimento} \textbf{A106}, 843 (1993).
\bibitem{AMS2} J. Sandweiss, {\em J. Phys.} \textbf{G30}, S51 (2004).
\bibitem{Madsen05} J. Madsen, {\em Phys. Rev.} \textbf{D71}, 014026 (2005). 
\bibitem{WilkF} G. Wilk and Z. W\l odarczyk, Proc. $23^{th}$ Int. Symp. Multiparticle Dynamics (2003), hep-ph/0210203; M. Rybczy\'nski et al., {\em Nucl. Phys.} (Proc. Suppl.) \textbf{B151}, 341 (2006); {\em Il Nuovo Cimento} \textbf{C24}, 645 (2001). 
\bibitem{ams} E. Finch,  {\em J. Phys.} \textbf{G32}, S251 (2006).
\bibitem{Arafune} J. Arafune et al., {\em Phys. Rev.}  \textbf {D62}, 105013 (2000).
\bibitem{SK2} Y. Takenaga et al., {\em Phys. Lett.} \textbf{B647}, 18 (2007), hep-ex/0608057. 
\bibitem{dama} F. Cappella et al., {\em Eur. Phys. J. direct} {\bf C14}, 1 (2002).
\bibitem{icecube} D. Hardtke et al., {\em Proc. International Workshop on Exotic Physics with Neutrino Telescopes}, 89 (2006).
\bibitem{pavalas} G.E. P\v{a}v\v{a}la\c{s}, {\em AIP Conf. Proc.} {\bf1304}, 454 (2010), arXiv:1010.2071 [astro-ph.HE].

\end{thebibliography}

\end{document}